\input{aipcheck}

\documentclass[
    ,final            
  ]
  {aipproc}

\layoutstyle{6x9}

\begin{document}

\title{Gamma-Ray Bursts in the SALT/Swift Era: GRB/SN Connection}


\author{Krzysztof Z. Stanek}{
  address={{\tt kstanek@cfa.harvard.edu}\\
Harvard-Smithsonian Center for Astrophysics\\
60 Garden St., MS20\\
Cambridge, MA 02138}
}

\begin{abstract}
Invited talk at the First International Workshop on ``Stellar
Astrophysics with the World Largest Telescopes'', Toru\'n, Poland,
7-10 September 2004. I discuss the Gamma Ray Burst research in the era
of SALT and Swift, concentrating on the GRB/SN connection.
\end{abstract}

\maketitle

\section{Introduction}

The mechanism that produces gamma-ray bursts (GRBs) has been the
subject of considerable speculation during the three decades since
their discovery (Klebesadel, Strong \& Olson~1973).  The CGRO and
BATSE observations demonstrated that GRBs are isotropically
distributed on the sky (Meegan et al.~1992), which could have been
explained with either a Galactic (Lamb 1995) or a cosmological
(Paczy\'nski 1995) spatial distribution.  The BeppoSAX satellite
(Boella et al.~1997) contributed the breakthrough in this field by
providing rapid and accurate localizations of X-ray afterglows of
GRBs.  Such precision allowed for quick optical (Groot et al.~1997;
van Paradijs et al.~1997) and radio (Frail et al.~1997)
identifications of transients associated with individual GRBs.  The
identification of the optical transient (OT) associated with
GRB\,970508 led to the first optical spectroscopic redshift
determination for a GRB, placing it at $z \geq 0.835$ (Metzger et
al.~1997), and thus firmly at a cosmological distance.  However,
despite the great progress made in GRB science since 1997, many
problems remain open. I discuss some of them, concentrating on the
GRB/SN connection. I also discuss some of the exciting GRB science
that could be done with the SALT telescope.

\section{GRB Science with Large Telescopes}

Large telescopes played a very significant role in the GRB science
since the discovery of first GRB afterglows in 1997.  Obviously, large
telescopes can do things that smaller telescopes cannot do, such as
obtaining spectroscopy of very faint sources (such as faint afterglows
or GRB host galaxies) or high-resolution spectroscopy of fairly faint
objects.  Perhaps more importantly, given the unpredictable timing of
GRBs, large telescopes allow us to obtain very significant results in
{\em relatively short} telescope time, defined as ``how much telescope
time can I ask for from a random (even if friendly) observer on a
remote telescope who has her/his own important program to do, before
he/she gets annoyed with me''?  It is my experience that up to one
hour per night is doable. It is important to realize that most large
telescope users do not sit around waiting for a GRB worker to call
them and tell them what to do, so common sense should be
applied. Common Polish heritage and passion for pierogies helps in
some cases.  Of course, with a queue-scheduled telescope like SALT,
the life of a GRB researcher becomes a bit easier, but not boring.

Below I list several areas of important GRB science explored with the
largest telescopes:

\begin{itemize}

\item Redshifts of the GRBs: This can be measured via absorption lines
in an intervening galaxy or galaxies (as was done for the first GRB
redshift by Metzger et al.~1997 using Keck 10-m telescope), via
emission lines of the host galaxy superimposed on the featureless
power-law spectrum of the afterglow (as was done for example for
GRB\,011121 by Garnavich et al.~2003a using Magellan 6.5-m telescope),
or via absorption lines in the spectrum of the host galaxy. For a
review on the hosts of GRBs, their redshifts and properties see
Djorgovski et al.~(2003) The first method, in principle, does not
require large telescopes, given a bright enough afterglow and the
presence of absorption lines (indeed, Jha et al. 2001 were able to
measure three different absorption systems in the afterglow of
GRB\,010222 using our FLWO 1.5-m telescope!), but large aperture
helps, given that most afterglows decay quickly.

\item High-resolution spectroscopy of GRB afterglows: High-resolution 
spectra have now been obtained for several afterglows (see for example
Schaefer et~al.~2003). In a number of cases a complicated line
structure with multiple components has been observed (see for example
a nice paper by Mirabal et al.~2003 on ``GRB 021004: A Possible Shell
Nebula around a Wolf-Rayet Star Gamma-Ray Burst Progenitor''), allowing
us to probe the enviroment of the GRB.

\item Optical polarization of the GRB afterglows: This is truly the
domain of the largest telescopes, given the photon-starved nature of
polarization research (i.e. trying to measure a very small effect in
the light of fairly faint objects). Despite this challenge,
polarization has been now successfully measured in a number of
afterglows.  The first detection was accomplished using imaging
polarimetry of a fairly bright afterglow of GRB\,990510 with the 8-m
VLT (Covino et al. 1999; Wijers et al. 1999).  Spectropolarimetry has
been obtained for several afterglows now as well, starting with
GRB\,020813 (Barth et al. 2003), using Keck 10-m telescope. The level
of optical polarization is usually low, 1-2\%, except in the case of
GRB\,020405, where 9.9\% polarization has been measured 1.3 days after
the burst (Bersier et al. 2003a) with the 6.5-m MMT.

\end{itemize}

The above was not intended as a review, but rather as a brief
illustration of some of the many interesting results obtained for the
GRBs with the largest telescopes. Needless to say, many exciting
results for the GRBs were obtained with mid-sized or in some cases
truly small telescopes (9th magnitude optical flash from the
GRB\,990123, observed with a 10-cm robotic ROTSE telescope, comes to
mind: Akerlof et al. 1999). For an interested reader, there are many
excellent reviews out there to learn more about GRBs. I found the
recent review by Piran (2004) to be very useful.

\section{Gamma-Ray Burst/Supernova Connection}

\begin{figure}
  \includegraphics[height=.65\textheight]{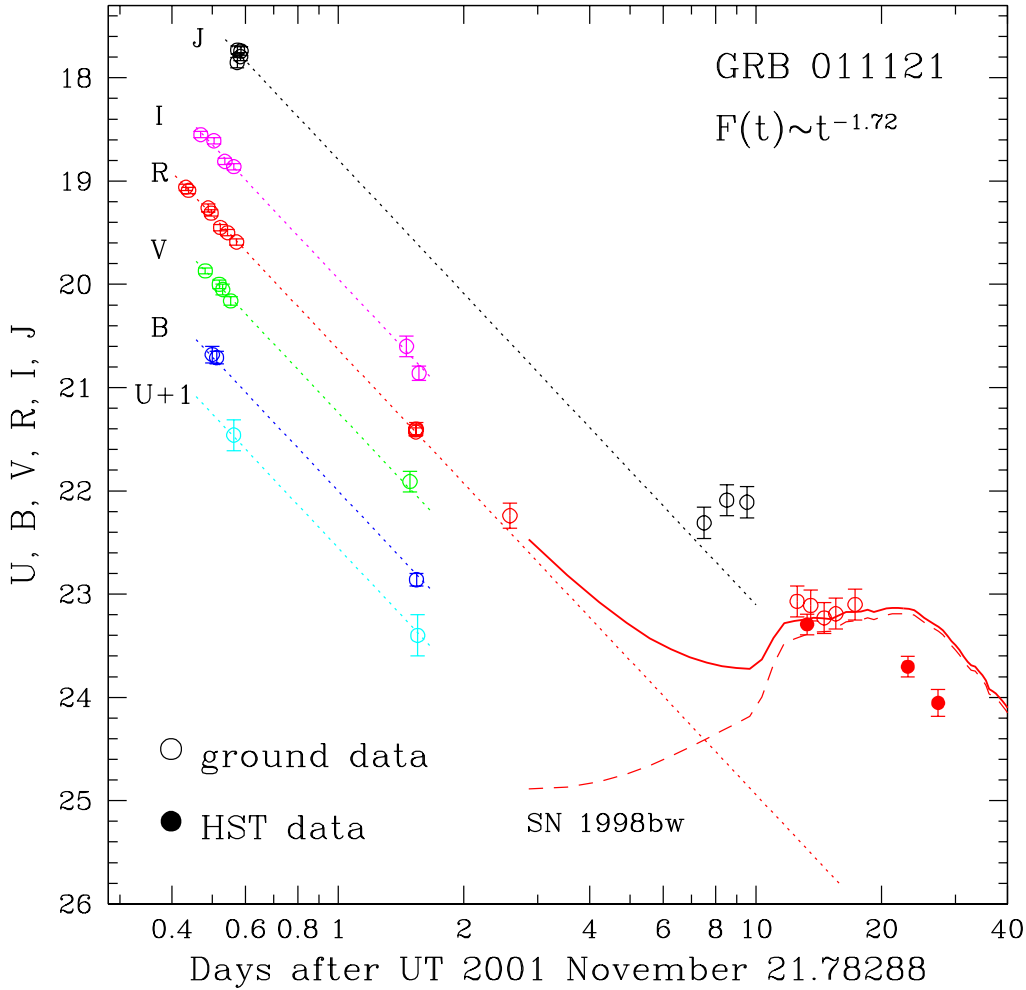} \caption{$UBVRIJ$\/
  light curves of GRB\,011121. Included are three {\em HST}\/ F702W
  epochs converted to the standard $R$-band.  Dotted lines show the OT
  power-law decay and the dashed line is the light curve of hypernova
  SN 1998bw redshifted to $z=0.36$, converted to the $R$-band,
  corrected for extinction and scaled by 0.1 mag.  The solid line
  shows the combination of the OA and SN~1998bw (from Garnavich et
  al. 2003a).}
\end{figure}

The measured redshift of a typical GRB is $z \approx 1$, implying that
a supernova component underlying an optical afterglow would be
difficult to detect.  At $z \approx 1$, even a bright core-collapse
event would peak at $R > 23$ mag.  Nevertheless, late-time deviations
from the power-law decline typically observed for optical afterglows
have been seen and these bumps in the light curves have been
interpreted as evidence for supernovae (for a recent summary, see
Bloom 2003).  GRB\,980425 was likely associated with ``hypernova''
1998bw (Galama et al.~1998), but the isotropic energy of that burst
was 10$^{-3}$ to 10$^{-4}$ times weaker than classical cosmological
GRBs which placed it in a unique class. Before March 2003, the best
evidence that classical, long-duration gamma-ray bursts are generated
by core-collapse supernovae was provided by GRB\,011121.  It was at $z
= 0.36$, so the supernova component would have been relatively bright.
A bump in the light curve was observed both from the ground and with
\emph{HST} (Garnavich et al.~2003; Bloom et al.~2002).  The color 
changes in the light curve of GRB\,011121 were also consistent with a
supernova (designated SN~2001ke), but a spectrum obtained by Garnavich
et al. (2003) during the time that the bump was apparent did not show
any features that could be definitively identified as originating from
a supernova.

\begin{figure}
  \includegraphics[height=.7\textheight]{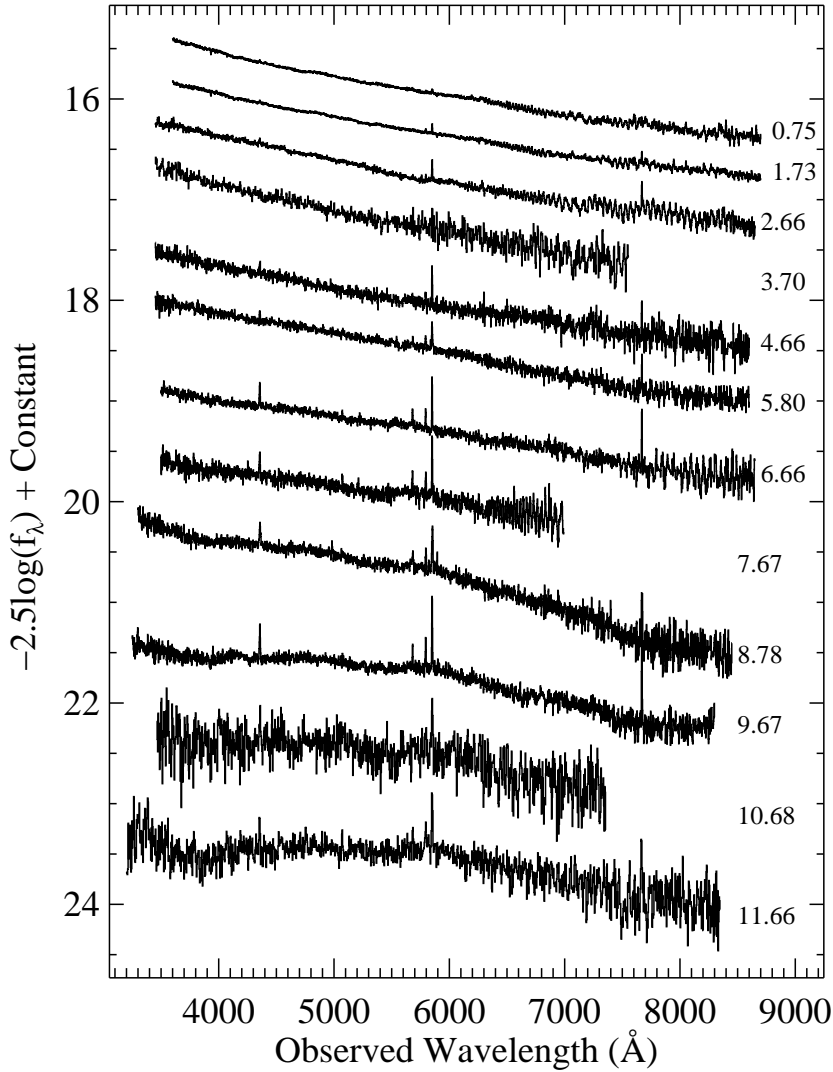} 
  \caption{Evolution of the GRB\,030329/SN~2003dh spectrum, from
  March~30.23 UT (0.75 days after the burst), to April 10.14 UT (11.66
  days after the burst). The early spectra consist of a power-law
  continuum with narrow emission lines originating from {\sc{H II}}
  regions in the host galaxy at $z = 0.1685$. Spectra taken after
  $\Delta T=6.66$ days show the development of broad peaks
  characteristic of a supernova (from Matheson et al.~2003b).}
\end{figure}

\begin{figure}
  \includegraphics[height=.65\textheight]{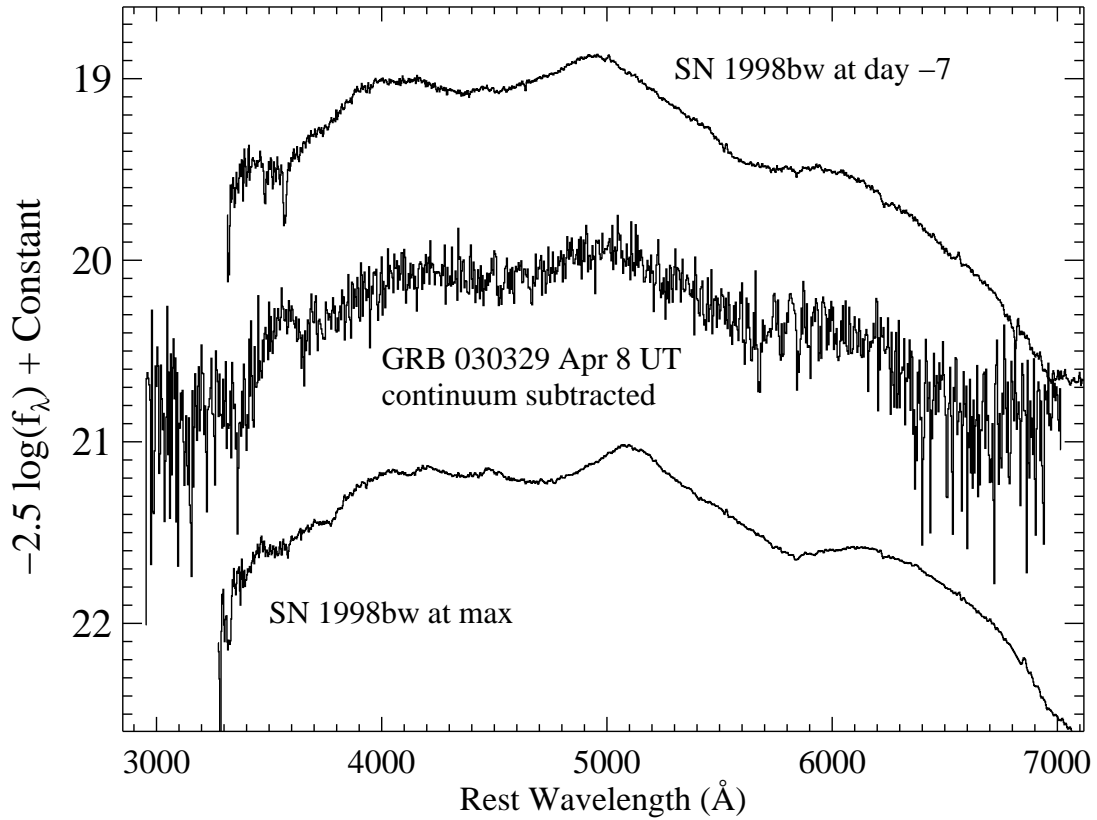}
  \caption{MMT spectrum of GRB\,030329 from April~8 with the smoothed
   MMT spectrum of April~1 scaled and subtracted.  The residual
   spectrum shows broad bumps at approximately 5000\AA\ and 4200\AA\
   (rest frame), which is similar to the spectrum of the peculiar type
   Ic SN\,1998bw a week before maximum light. The match is not as good
   for SN\,1998bw at maximum light (from Stanek et al. 2003).}
\end{figure}

\begin{figure}
  \includegraphics[height=.7\textheight]{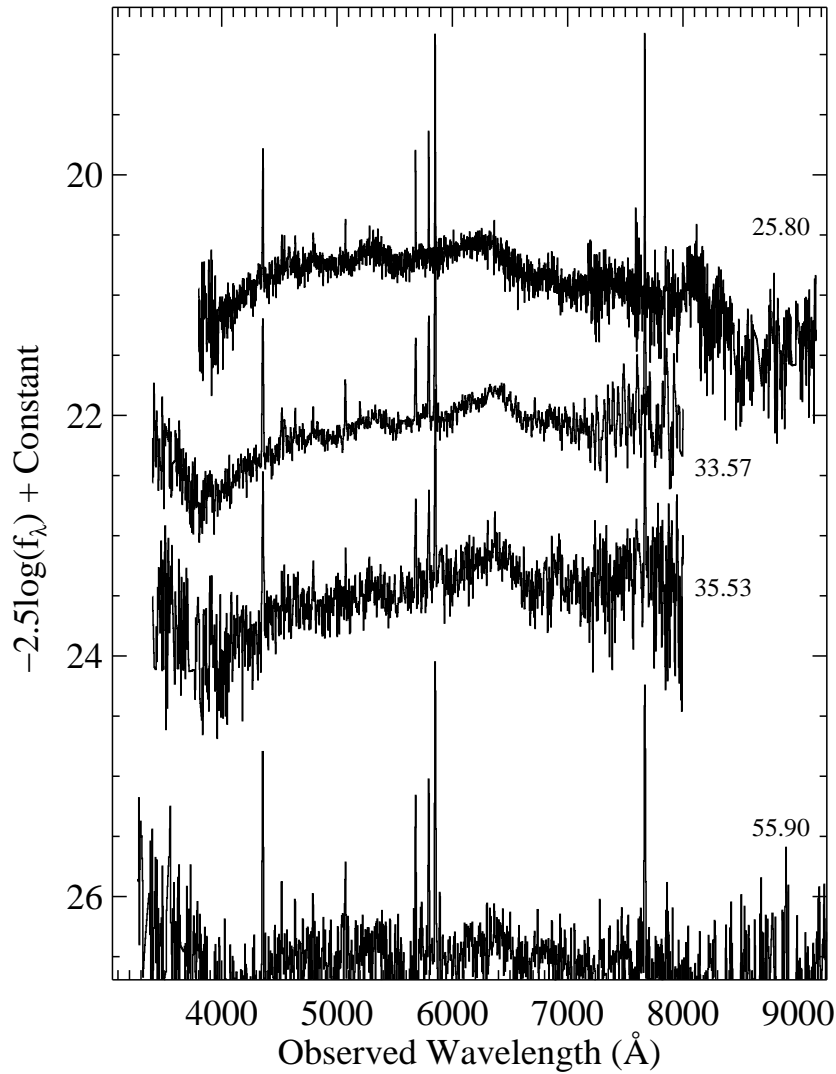} 
  \caption{Evolution of the GRB\,030329/SN~2003dh spectrum, from April
  24.28 UT (25.8 days after the burst), to May 24.38 (55.9 days after
  the burst).  The power-law contribution decreases and the spectra
  become more red as the SN component begins to dominate. The upturn
  at blue wavelengths may still be the power law.  The broad features
  of a supernova are readily apparent, and the overall spectrum
  continues to resemble that of SN~1998bw several days after maximum
  (from Matheson et al. 2003b). }
\end{figure}

\begin{figure}
  \includegraphics[height=.55\textheight]{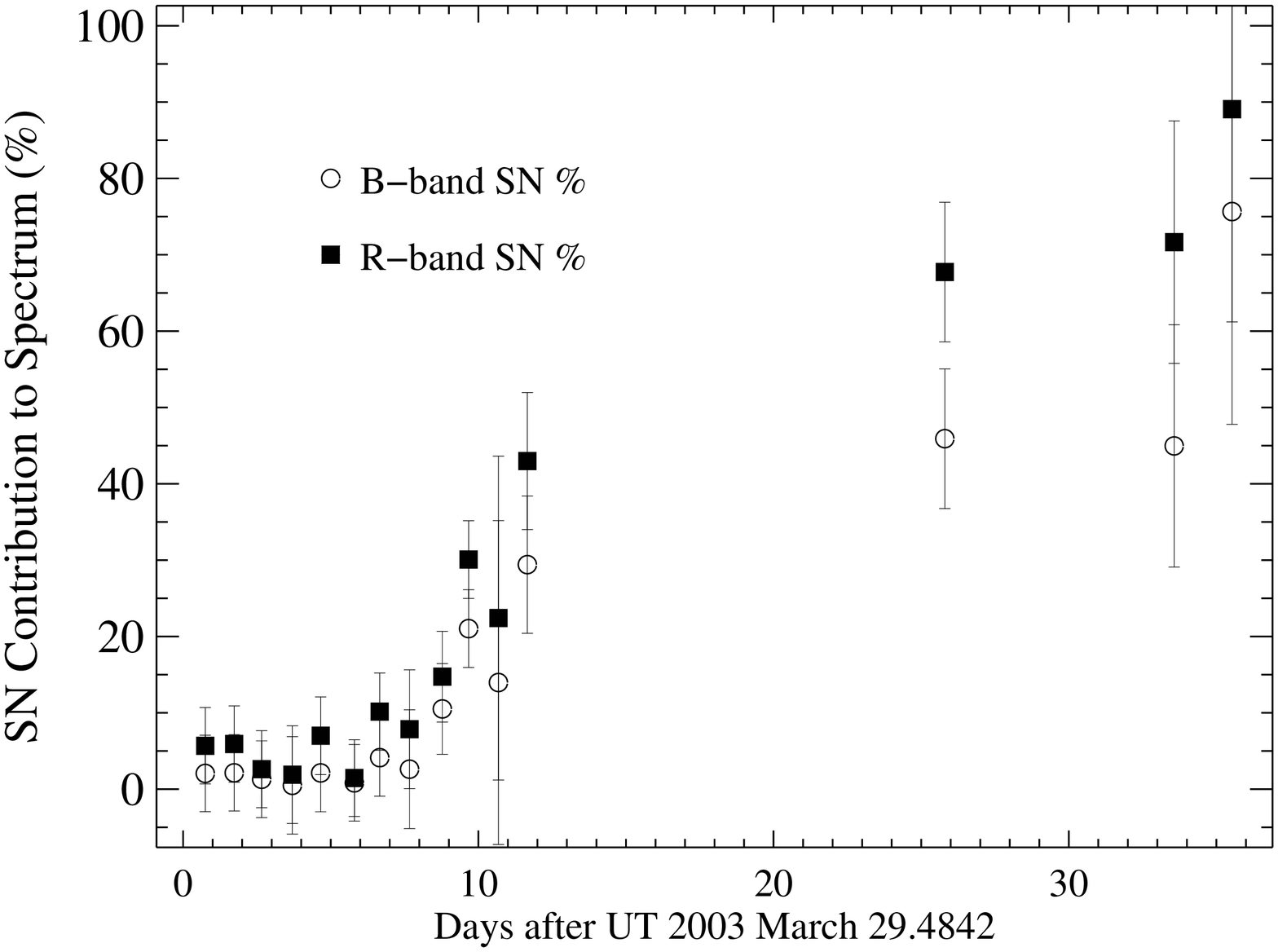} 
  \caption{Relative contribution of a supernova spectrum to the
  GRB\,030329/SN~2003dh afterglow as a function of time in the $B$
  (\emph{open circles}) and $R$ (\emph{filled squares}) bands.  We
  derived a best fit to the afterglow spectrum at each epoch with the
  fiducial power-law continuum and the closest match from our set of
  peculiar SNe Ic.  We then synthesize the relative $B$-band and
  $R$-band contributions.  There is some scatter for the early epochs
  due to noise in the spectra, but a clear deviation is evident
  starting at $\Delta T=7.67$ days, with a subsequent rapid increase
  in the fraction of the overall spectrum contributed by the SN.
  Errors are estimated from the scatter when the SN component is close
  to zero ($\Delta T < 6$ days) and from the scale of the error in the
  least-squares minimization (from Matheson et al. 2003b).  }
\end{figure}

While SN~1998bw was a strong hint of the GRB-SN association, no
optical afterglow was observed.  Without that direct association, the
link between GRBs and SNe was still in question.  The `monster burst'
of 2003, GRB\,030329 provided that link.  The burst was extremely
bright in gamma-rays, implying that it was relatively close, as was
quickly confirmed by the VLT spectroscopy of the afterglow, which
yielded $z=0.168$ (Greiner et~al. 2003).  As the afterglow faded,
subtle features appeared in the normally flat power-law spectrum of
the afterglow.  By subtracting a continuum based upon the early shape
of the spectrum, this structure was revealed as the spectrum of an
unusual Type Ic SN similar to SN~1998bw, designated SN~2003dh
(Matheson et al. 2003a; Garnavich et~al. 2003b; Stanek et~al. 2003).
Within a few days, the SN became the dominant component in the
spectrum (Stanek et~al. 2003; Kawabata et~al. 2003; Hjorth et
al. 2003w).

Using the early power-law continuum spectrum as a model, one could
decompose the observed spectra at later times into two separate
components: GRB afterglow and SN spectrum.  Using a least-squares
technique, the best match for the SN among the low-redshift sample was
SN~1998bw (Matheson et~al. 2003b).  In fact, taking into account
cosmological time dilation, the spectroscopic evolution of SN~2003dh
almost exactly matched SN~1998bw.  Models of these spectra are
presented by Mazzali et al. (2003).

An important point about the appearance of the SN was that the light
curve did not show the bump that is characteristic of a rebrightening
caused by the SN (see Matheson et~al. 2003b and Lipkin et~al. 2004 for
a discussion of the light curve, which actually showed many bumps).
Without the spectroscopic confirmation, the presence of SN in
GRB\,030329 would still be argued about.

Nebular-phase spectra of SN~2003dh show a spectrum similar to a
typical Type Ic SN.  Kosugi et~al. (2004) present a spectrum at an age
of $\sim$3 months.  A spectrum obtained with the Keck telescope by
Filippenko, Chornock, \& Foley (2004) in December of 2003 is much like
a normal Type Ic SN (Bersier et~al. 2005, in preparation).

Following the discovery of SN~2003dh, reexamination of spectra of an
earlier burst yielded some evidence for a SN component.  Della Valle
et~al. (2003) found that a spectrum of the very faint afterglow of
GRB\,021211 had structure similar to an SN.  In this case, however,
the SN did not match SN~1998bw or any other peculiar Type Ic SN, but
it was most similar to SN~1994I, a relatively normal Type Ic.

Another example of the GRB/SN connection came with GRB\,031203.
Despite high foreground reddening, spectroscopy with the VLT revealed
an SN component, designated SN~2003lw (Malesani et~al. 2004).  For
this SN, SN~1998bw was again a good match.  Of the four SNe with clear
GRB associations, three show remarkably similar spectra.

\begin{figure}
  \includegraphics[height=.6\textheight]{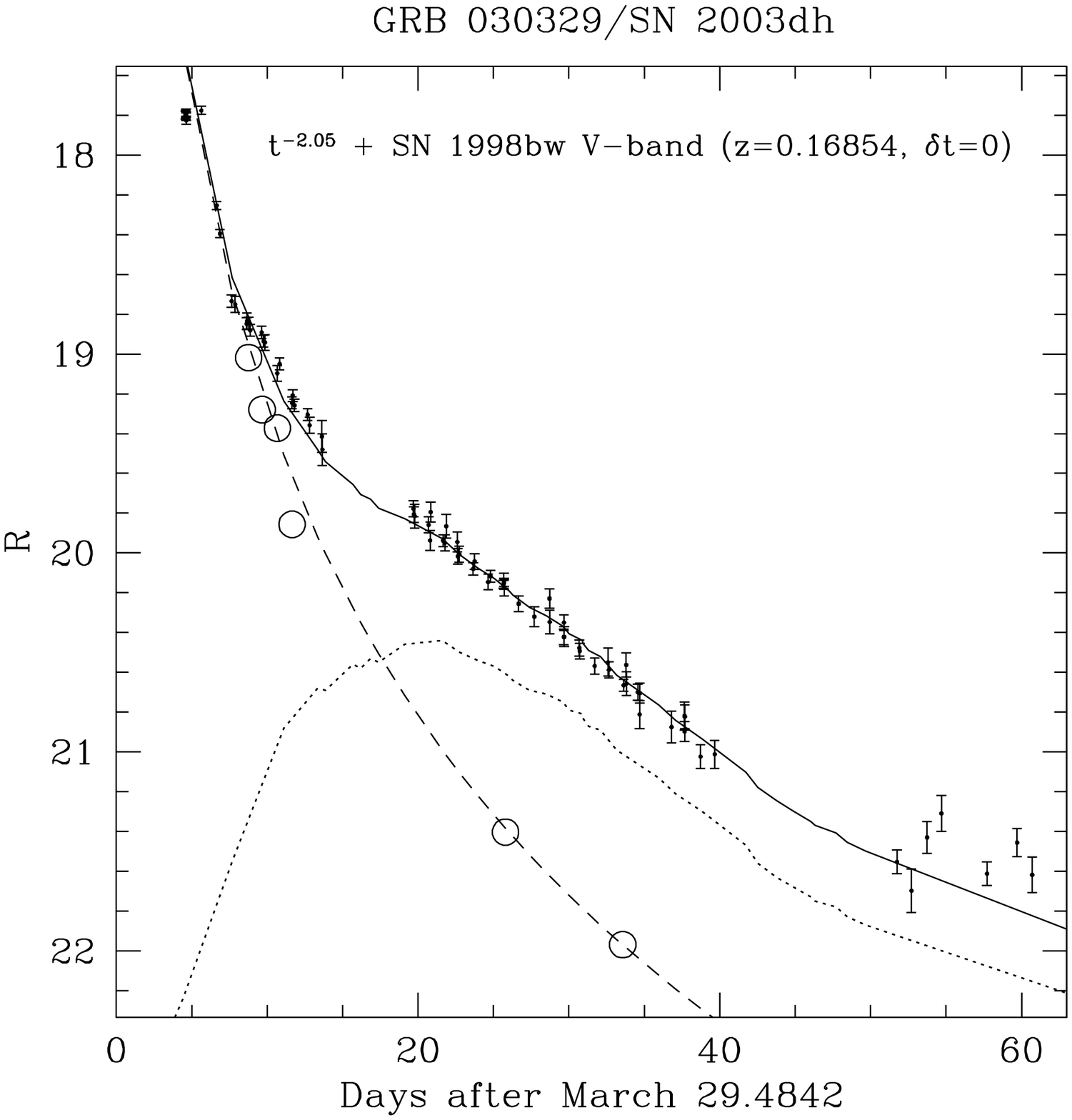}
   \caption{Decomposition of the GRB\,030329 OT $R$-band light curve
   into the supernova (\emph{dotted line}) and the power-law continuum
   (\emph{dashed line}).  As the light curve model for the supernova,
   we took the $V$-band light curve of SN~1998bw stretched by
   $(1+z)=1.1685$.  The resulting supernova light curve peaks at
   $m_R=20.4$ (from Matheson et al. 2003b).}
\end{figure}

\section{Future directions of the GRB/SN connection}

There is now indisputable evidence that some classical (long
duration/soft spectrum) GRBs come from core-collapse supernovae. But
there are only two classical bursts (GRB\,030329 and GRB\,031203)
where the progenitor supernovae have been definitively classified.
SN~2003dh and SN~2003lw were type~Ic events similar to the progenitor
of the underluminous GRB\,980425. Some theorists have concluded that
all supernovae that produce GRB are type~Ic. The bias toward type~Ic
is partly due to the perceived difficulty in getting the jet to escape
from a star with a massive envelope.  But this prejudice may not be
justified; after all, it was recently believed that no supernova could
possibly be a GRB source because of the large baryon content of
supernovae. I feel that the range of SN types that are responsible for
GRBs is an unsolved observational problem.

There are hints of a diversity in the supernova progenitors of
GRB. SN1997cy was a powerful type~II supernova that may have been
associated with a classical GRB (Germany et al. 2000), though
connection to a particular GRB has not been strongly established. A
better indication comes from GRB\,011121 and its associated supernova
2001ke. Garnavich et al. (2003a) found the color near maximum of
SN~2001ke was very blue which is more consistent with a type~II event
than a type~Ic. Unfortunately, there was no spectrum of the event with
sufficient quality to establish the supernova type.

The small number of supernovae directly associated with GRB also
leaves the question of the energetics unanswered.  The supernovae that
produced GRB\,980425 and 030329 were two of the most energetic
core-collapse events recorded.  But the supernova that produced
GRB\,011121 was somewhat fainter and faded quickly. What is the range
of supernova luminosities?  Is there a connection between the energy
of core collapse and the GRB properties? Frail et al. (2001) corrected
the observed GRB energy for the beam opening angle and found that most
bursts produce a total energy of $10^{51}$~erg.  While this is a very
large amount of energy, it is still an order of magnitude less than
the total energy in a typical core-collapse supernova.  Clearly, there
is sufficient energy in a core-collapse to allow a wide range of
supernovae to produce high-energy beams. The greatly improved
sensitivity of {\em Swift}\/ will allow low-energy bursts to be
detectable out to a large volume of space. This will be an important
test of the single-energy reservoir theory as well as the ability of
supernovae with a range of properties to create GRB.

The few supernovae associated with GRBs also lead theorists to
conclude that GRB are produced promptly by the core-collapse of
massive stars.  But this could be an observational bias.  It is
possible that GRBs occur within a wide range of times following
core-collapse, and we only detected those few bursts that occurred
less than a few days after the supernova. In support of this
possibility, a number of GRB have not been seen with accompanying
supernovae (e.g. Price et al.~2003).  The `supranova' model predicts
that the final collapse of a neutron star into a black hole can be
delayed by seconds to years after the initial core-collapse. While it
would be difficult to show a supernova went off years before a burst
using optical wavelengths, supernovae that exploded weeks to a month
before the GRB could be detected with careful observation.

Obtaining magnitudes, colors and spectra of more GRB supernovae is
clearly a top priority in understanding the origin of long/soft
bursts.  Since the typical GRB has a redshift of $z\sim 1$, this work
requires large telescopes such as SALT.

\section{A Modest Proposal for SALT in the Swift Era}

The Swift satellite (Gehrels et al.~2004) will push the GRB research
into an even higher gear, with $>100$ GRBs per year rapidly and
accurately localized. SALT is very well positioned to take full
advantage of Swift, due to a number of factors:

\begin{itemize}

\item Unique position on the globe;

\item Queue schedule;

\item Stable instrumentation for several years;

\end{itemize}

For the description of SALT and its capabilities, see Buckley 2004
(this volume).  The first item on the list is obvious, i.e. despite
limitations on pointing, due to its location in South Africa SALT will
be able to access some of the GRBs not possible to observe from the
North or to observe them earlier than possible from Chile, where other
large southern telescopes are located. Queue scheduling is even more
important, allowing certain programs to be executed which would not be
possible otherwise, for example to observe an object each night for a
month or a year. Stability of the instrumentation
(PFIS\footnote{\url{http://www.sal.wisc.edu/pfis/}}---The Prime Focus
Imaging Spectrograph---will be always available for the first several
years), combined with the queue schedule, will allow one to obtain
very uniform data sets over long periods of time, which is very hard
to do on other large telescopes.

Here are several GRB projects to consider over the first several years
of the SALT operation:

\begin{itemize}

\item GRBs-SNe connection: observe every $z<0.5$ GRB afterglow (several
      a year) at 15-20 epochs during the first month after the burst
      and during 5-10 epochs later. This will answer the questions
      posed above: what are the GRB progenitors, do they form a
      uniform class?

\item Physics of the afterglow: obtain polarization light curves
      of several bright afterglows per year for 3-7 days each.  This
      will be important to understand the physics of the GRB emission
      and to test the jet model of the afterglow.

\item Very short timescale variability of GRB afterglows. We have
      reported short-timescale afterglow variability in two cases
      (GRB\,011211: Holland et al.~2002; GRB\,021004: Bersier at
      al.~2003b). In some cases the afterglows are very smooth
      (GRB\,990510: Stanek et al. 1999; short-timescale variability
      $<$0.5\% for GRB\,020813: Laursen \& Stanek 2003). Again, the
      ability of SALT to perform very rapid photometry (Charles 2004,
      this volume) is unique among the largest telescopes.

\end{itemize}

Many other projects can be proposed, here I tried to concentrate on
areas where the strengths of SALT would be well utilized. Earlier
during the conference, we were encouraged to think about truly large
projects (``what would you do with infinite time on SALT'') to be done
with SALT during the first few years. Being a reasonable person, I
decided not to use ``infinite time'', just 10\% of it, to spend one
hour {\em every night} to obtain various kinds of data (depending on
the brightness) for any GRB afterglow that is observable from SALT
that night.  Such a project would be a major undertaking, but with the
expected Swift GRB rate the sample of well observed afterglows would
easily {\em double} in just several months, and there would be many
surprises and new exciting results.

\begin{theacknowledgments}

I wish to thank my colleagues and friends from the CfA/Notre Dame GRB
group, Peter Garnavich, David Bersier, Stephen Holland, Saurabh Jha
and Tom Matheson for their excellent collaboration.  I also want to
thank all the astronomers (about 100 of them so far, too many to be
named here) who devoted their telescope time and effort to make this
research possible. Some of them collaborated with us on multiple
bursts, and I hope we have many future bursts to observe.  I thank Tom
Matheson, Andy Szentgyorgyi and Joel Hartman for useful comments on
this manuscript. I would also like to thank Joanna Mikolajewska and
other organizers of the First International Workshop on ``Stellar
Astrophysics with the World Largest Telescopes'' for their invitation
and a very enjoyable conference.

\end{theacknowledgments}


\end{document}